\documentclass[prb,aps,showpacs,floats,twocolumn]{revtex4}
\usepackage{graphicx}

\begin{document}

\title{To probe quantum criticality with scanning tunneling spectroscopy}

\author{Minh-Tien Tran$^{1,2}$ and Ki-Seok Kim$^1$}
\affiliation{
$^1$Asia Pacific Center for Theoretical Physics, POSTECH, Pohang, Gyeongbuk 790-784, Republic of Korea \\
$^2$Institute of Physics, Vietnamese Academy of Science and Technology, P.O.Box 429, 10000 Hanoi, Vietnam}

\begin{abstract}
We investigate the role of quantum coherence in tunneling
conductance, where quantum criticality turns out to suppress Fano
resonance. Based on the nonequilibrium noncrossing approximation,
we show that the linear tunneling conductance exhibits weak Fano
line-shape with sharp cusp at zero energy in the multichannel
Kondo effect, resulting from incoherence associated with quantum
criticality of impurity dynamics. In particular, shift of the peak
position in the Fano resonance is predicted not to occur for the
multichannel Kondo effect, distinguished from the Fermi liquid
theory in the single channel Kondo effect.
\end{abstract}


\maketitle

\section{Introduction}

Recently, scanning tunneling microscopy (STM) has been utilized
extensively to probe the electronic structure of materials with
atomic scale spatial resolution. It was found that formation of
the Kondo resonance gives rise to an asymmetric line-shape in the
tunneling conductance through the STM tip close to a magnetic
adatom on a metallic surface,\cite{Madhavan,Li,Manoharan} the
origin of which is an effect of interference between the direct
tip-host tunneling and indirect tip-adatom-host one that
resembles so called Fano resonance.\cite{Fano} This Fano-Kondo
effect has been discussed in detail.\cite{Ujsaghy,Schiller,Plihal}
A similar Fano-Kondo effect was also investigated in the electron
transport through a quantum dot embedded in a closed Aharonov-Bohm
interferometer.\cite{Gores,Hofstetter}

Mechanism of Fano resonance implies that quantum coherence of
impurity dynamics plays an important role for the line-shape,
where the coherence time scale is estimated as $\sim 1/T_{K}$ with
the Kondo temperature $T_{K}$. An interesting question is what
happens in the Fano line-shape if impurity dynamics becomes
incoherent. Such a situation is realized in the multichannel Kondo
system, where screening of a local moment by conduction electrons
is overcompensated to drive the local Fermi liquid state into a
non-Fermi liquid critical state, first suggested by Nozieres and
Blandin in the multichannel Kondo model.\cite{Nozieres}

Multichannel Kondo impurity systems have been studied both
experimentally and theoretically with considerable interests.
Recently, the multichannel Kondo model was realized artificially
in quantum dots.\cite{Oreg} The multichannel Kondo effect was also
claimed to occur in the quadrupolar Kondo effect\cite{Seaman} and
in metal point contacts.\cite{Ralph} In the theoretical respect
the multichannel Kondo model has been studied in a variety of
controlled techniques.\cite{CoxZawadowski} The conformal field
theory approach provides exact results of the non-Fermi liquid
fixed point,\cite{Affleck} and the noncrossing approximation
(NCA), exact in the limit of large number of spin flavors and
charge channels, also gives practically sensible
results,\cite{Cox,Parcollet} where universal power-law scaling is
found in physical responses. Such power-law physics distinguishes
the critical non-Fermi liquid state of the multichannel Kondo
impurity from the local Fermi liquid state of the single-channel
Kondo impurity.

In this paper we study the Fano-Kondo effect by the tunneling
current which flows from a single-channel STM tip to a
multichannel Kondo impurity host. Instead of the standard
Fano-Kondo resonance in the tunneling conductance, one may expect
different pronounced features due to interference between the Fano
resonance of the tunneling current and the overcompensated
screening of an impurity. Employing the Keldysh nonequilibrium
formalism,\cite{Keldysh,Meir} we derive the tunneling current and
its conductance, where the tunneling current solely depends on the
Green function of an impurity. We calculate the linear conductance
profile analytically at zero temperature, based on the
nonequilibrium NCA\cite{Wingreen,Hettler} to obtain nonequilibrium
Green functions of the impurity. A power-law line-shape in the
tunneling conductance clearly shows the overscreening effect of an
impurity. Such an effect of incoherence leads the Fano resonance
suppressed, and its asymmetric feature becomes considerably weak.
First of all, shift of the peak position in the Fano resonance
turns out not to occur in the multichannel Kondo effect. These
features are argued to be quite general, distinguishing the
non-Fermi liquid phase from the Fermi-liquid state in STM.

The plan of the present paper is as follows. In Sec.~II we present
our model on an STM setup and derivation for the tunneling
current. We introduce the nonequilibrium NCA in Sec.~III. The
conductance profile is analyzed at zero temperature in Sec.~IV.
Finally, the conclusion is presented in Sec.~V.

\section{tunneling conductance}

\subsection{Model}

\begin{figure}[t]
\includegraphics[width=0.4\textwidth]{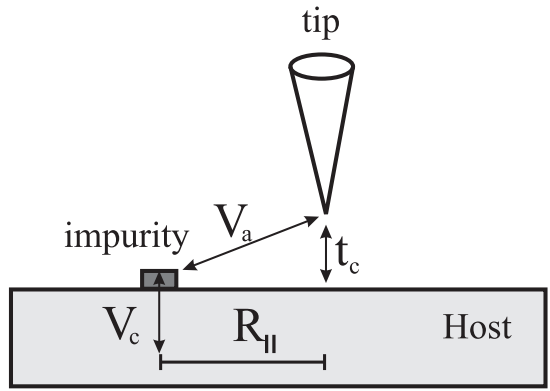}
\caption{Scanning tunneling microscope (STM) device with a tip placed
closely to a Kondo impurity on the surface of a normal metal (host). In the
host the impurity is hybridized with the metal conduction band through
coupling $V_c$. The STM tip couples to the impurity via hopping $V_a$
and to the local conduction electrons of the host via hopping $t_c$. }
\label{fig1}
\end{figure}

The system under consideration is shown schematically in
Fig.~\ref{fig1}. It consists of a multichannel Kondo impurity host
and a single-channel STM tip, placed directly above the host
surface. The multichannel Kondo impurity host is modelled by a
multichannel Anderson model in the slave-boson representation,
which explicitly separates spin and channel excitations.\cite{Cox}
The STM tip couples separately to the impurity and to the local
conduction electrons of the host.

The Hamiltonian of the system takes the form
\begin{eqnarray}
H &=& H_{\rm host} + H_{\rm tip} + H_{\rm tunn} \label{ham}\\
H_{\rm host} &=& \sum_{\mathbf{k},\sigma,\tau} \varepsilon_{\mathbf{k}}
c^{\dagger}_{\mathbf{k}\sigma\tau} c^{\null}_{\mathbf{k}\sigma\tau} +
\varepsilon_{f} \sum_{\sigma} f^{\dagger}_{\sigma} f^{\null}_{\sigma}  \nonumber \\
&& +
V_{c} \sum_{\mathbf{k},\sigma,\tau} f^{\dagger}_{\sigma} b_{\bar{\tau}} c^{\null}_{\mathbf{k}\sigma\tau}
+ {\rm h.c.} , \label{ham2} \\
H_{\rm tip} &=& \sum_{\mathbf{k},\sigma} ( E_{\mathbf{k}} - eV)
a^{\dagger}_{\mathbf{k}\sigma} a^{\null}_{\mathbf{k}\sigma} , \label{ham3} \\
H_{\rm tunn} &=& \sum_{\mathbf{k},\sigma,\tau} V_{a}(\mathbf{R}_t)
f^{\dagger}_{\sigma} b_{\bar{\tau}} a^{\null}_{\mathbf{k}\sigma} +  {\rm h.c.}  \nonumber \\
&& + \sum_{\mathbf{k},\mathbf{p},\sigma,\tau} t_c(\mathbf{k},\mathbf{R}_{\|}) c^{\dagger}_{\mathbf{k}\sigma\tau}
a^{\null}_{\mathbf{p}\sigma}
+ {\rm h.c.} , \label{ham4}
\end{eqnarray}
where $c^{\dagger}_{\mathbf{k}\sigma\tau}$
($a_{\mathbf{k}\sigma}$) creates host (tip) conduction electrons
with wave vector $\mathbf{k}$, spin $\sigma$, and channel $\tau$.
The spin degeneracy is $N$ and
the number of channels is $M$.
In the slave boson representation the impurity creation operator
is given by $f^{\dagger}_{\sigma} b_{\bar{\tau}}$, where
$f_{\sigma}$ is a fermion operator and $b_{\bar{\tau}}$ is a boson
operator. The fermion, $f^{\dagger}_{\sigma}$, transforms
according to $SU(N)$ and creates a local spin excitation, whereas
the boson $b_{\bar{\tau}}$ transforms according to the conjugate
representation of $SU(M)$ and annihilates the channel quantum
number of the "vacuum" state produced by destroying a conduction
electron.\cite{Cox,Parcollet}  Completeness of
local states at the impurity site is represented by the constraint
$\sum_{\sigma} f^{\dagger}_{\sigma} f^{\null}_{\sigma} +
\sum_{\bar{\tau}} b^{\dagger}_{\bar{\tau}} b^{\null}_{\bar{\tau}}
= 1$, implemented as usual by introducing a Lagrange multiplier
$\lambda$. $\varepsilon_{\mathbf{k}}$ is the band dispersion of
conduction electrons in the host, and $\varepsilon_{f}$ is the
energy level of the Kondo impurity or the adatom at the host
surface. $V_{c}$ is the hybridization parameter of the impurity
and conduction electrons in the host. For simplicity, $V_c$ is
assumed to be constant. $E_{\mathbf{k}}$ is the band dispersion of
tip-conduction electrons, and $eV$ is an applied voltage bias
between the tip and host, which causes a weak electron current to
flow between them. We set the chemical potential of host
conduction electrons as our reference energy.

Couplings between the impurity and tip-conduction electrons are
represented by the hybridization parameter $V_a(\mathbf{R_t})$,
where $\mathbf{R}_{t}$ is the tip position. It decays with a
tip-to-impurity separation\cite{Plihal}
$$
V_a(\mathbf{R_t}) \approx V_a e^{ - \kappa |R_t|},
$$
where $\kappa$ is an effective decay constant evaluated for states
at the Fermi level of the tip. In this paper we consider $|R_{t}|
\ll 1/\kappa$, thus $V_a(\mathbf{R_t})$ is modelled as a constant.
Couplings between the tip and host conduction electrons are
represented by $t_c(\mathbf{k},\mathbf{R}_{\|})$, where
$\mathbf{R_{\|}}$ is a parallel distance between the tip and
impurity. For plane waves of conduction electrons, we have
$$
t_c(\mathbf{k},\mathbf{R}_{\|}) = t_{c} e^{- i \mathbf{k} \mathbf{R}_{\|} } .
$$
When the tip is placed directly on top of the impurity, we take
$t(\mathbf{k},\mathbf{R}_{\|}) = t_c$.

The host Hamiltonian of Eq.~(\ref{ham2}) is the multichannel
Anderson model. In the case of $M=N$ the impurity is completely
screened to form the Kondo singlet, resulting in the local Fermi
liquid. When the number of channels is larger than the spin
degeneracy ($M>N$), the impurity is overcompensated to give rise
to a non-Fermi liquid fixed point, which exhibits universal
power-law scaling.\cite{Affleck,Cox,Parcollet} The universal
scaling property lies at the heart of quantum critical phenomena
in a number of materials. In this respect the present study can be
said to probe non-Fermi liquid physics with STM. Unfortunately,
experimental realization of the multichannel Anderson model is
still problematic. Recently, the two-channel Kondo effect was
realized artificially in quantum dots.\cite{Oreg} If the STM tip
is applied to one of the leads, such non-Fermi liquid physics
would be observed.

\subsection{Tunneling current}

The electron current flowing between the tip and host is
calculated within the Keldysh nonequilibrium
formalism.\cite{Keldysh,Meir} The current from the tip to the host
is given by the time evolution of the occupation number for the
electrons in the tip
\begin{equation}
J_{t\rightarrow h}(t) = e \bigg\langle \frac{d N_{t}}{d t} \bigg\rangle =
- \frac{i e}{\hbar} \Big\langle \big[H(t), N_t(t)\big] \Big\rangle,
\end{equation}
where $N_t = \sum_{\mathbf{k}\sigma}
a^{\dagger}_{\mathbf{k}\sigma} a^{\null}_{\mathbf{k}\sigma}$. One
can express the current via nonequilibrium Green functions
\begin{eqnarray}
J_{t\rightarrow h}(t) &=& \frac{e}{\hbar}
\sum_{\mathbf{k},\sigma,\tau} V_{a}^{*} G^{<}_{d\sigma\tau,a\mathbf{k}\sigma}(t,t) \nonumber \\
&+& \frac{e}{\hbar}
\sum_{\mathbf{k},\mathbf{p},\sigma,\tau} t_{c}^{*} G^{<}_{c\mathbf{p}\sigma\tau,a\mathbf{k}\sigma}(t,t) +
{\rm h.c.},
\end{eqnarray}
where
\begin{eqnarray}
G^{<}_{d\sigma\tau,a\mathbf{k}\sigma}(t,t') &=& i \big\langle
a^{\dagger}_{\mathbf{k}\sigma}(t') d_{\sigma\tau}(t)
\big\rangle , \\
G^{<}_{c\mathbf{p}\sigma\tau,a\mathbf{k}\sigma}(t,t) &=&
i \big\langle
a^{\dagger}_{\mathbf{k}\sigma}(t') c_{\mathbf{p}\sigma\tau}(t)
\big\rangle ,
\end{eqnarray}
are lesser Green functions. Here we use the notation
$d_{\sigma\tau}=b^{\dagger}_{\bar{\tau}} f_{\sigma}$.

In the steady state nonequilibrium Green functions depend only on
$t-t'$, and the Fourier transformation results in
\begin{eqnarray}
J_{t\rightarrow h} &=& \frac{e}{\hbar}
\sum_{\sigma,\tau} \int d\omega \; V_{a}^{*} G^{<}_{d\sigma\tau,a\sigma}(\omega) \nonumber \\
&+& \frac{e}{\hbar}
\sum_{\sigma,\tau} \int d\omega \; t_{c}^{*} G^{<}_{c\sigma\tau,a\sigma}(\omega) +
{\rm h.c.},
\label{current}
\end{eqnarray}
where $G^{<}_{d\sigma\tau,a\sigma}(\omega)$,
$G^{<}_{c\sigma\tau,a\sigma}(\omega)$ are the Fourier
transformations of
$\sum_{\mathbf{k}}G^{<}_{d\sigma\tau,a\mathbf{k}\sigma}(t,t')$,
$\sum_{\mathbf{k},\mathbf{p}}G^{<}_{c\mathbf{p}\sigma\tau,a\mathbf{k}\sigma}(t,t')$,
respectively. These nonequilibrium Green functions can be
expressed via the impurity Green function based on the equation of
motion method. Detailed calculations are presented in Appendix A.

In a similar way we can calculate a current flowing from the host
to the tip
\begin{equation}
J_{h\rightarrow t}(t) = e \bigg\langle \frac{d N_{c}}{d t}
\bigg\rangle = - \frac{i e}{\hbar} \Big\langle \big[H(t),
N_c(t)\big] \Big\rangle   \label{currentht}
\end{equation}
with $N_c = \sum_{\mathbf{k}\sigma\tau}
c^{\dagger}_{\mathbf{k}\sigma\tau}
c^{\null}_{\mathbf{k}\sigma\tau}$, where its detailed expression
is given by Eq. (\ref{currenthta}) in Appendix A.

Calling $J_{h\rightarrow t}=-J_{t\rightarrow h}$ in the steady
state, the steady current can be rewritten in the form
\begin{equation}
J = y J_{t\rightarrow h} - (1-y) J_{h\rightarrow t} ,
\label{currenty}
\end{equation}
where $y$ is an arbitrary number. We choose $y$ such that the term
associated with the lesser Green function
$G^{<}_{d\sigma\tau,d\sigma\tau}(\omega)$ vanishes in the current
formula from Eq. (\ref{currentth}) and Eq. (\ref{currenthta}). As
a result, the electron current flowing between the tip and the
host reads
\begin{eqnarray}
J = \frac{e}{\hbar} \sum_{\sigma\tau} \int \frac{d\omega}{2\pi} T_{\rm tr}(\omega)
\big[f_a(\omega)-f_c(\omega) \big] ,
\label{currentfinal}
\end{eqnarray}
where
\begin{equation}
T_{\rm tr}(\omega) = T_0 + Q_{R}  {\rm Re} G^{R}_{d\sigma\tau,d\sigma\tau}(\omega)
+ Q_I  {\rm Im} G^{R}_{d\sigma\tau,d\sigma\tau}(\omega) ,
\label{trans}
\end{equation}
and $f_{a(c)}(\omega)$ is the Fermi-Dirac distribution function for tip (host)
conduction electrons.
$T_0$ and the coefficients $Q_{R(I)}$ are defined as
\begin{eqnarray*}
T_0 &=& \frac{4\gamma}{(1+M\gamma)^2} , \\
Q_{R} &=& 8 \frac{1-M\gamma}{(1+M\gamma)^3} \sqrt{\gamma \Gamma_a \Gamma_c} , \\
Q_{I} &=&
\frac{4}{(1+M\gamma)^3} \frac{1}{\Gamma_s}
(\Gamma_a+\gamma \Gamma_c)(\Gamma_c+M\gamma \Gamma_a)\\
&-& \frac{4(1-M\gamma)}{(1+M\gamma)^4} \frac{1}{\Gamma_s}
\Big[ (\Gamma_a-\gamma \Gamma_c)
\big(
M\gamma \Gamma_a + \\
&& \Gamma_c \big((1+M\gamma)(\gamma (M-1)+1)-\gamma \big)
\big)  + \\
&& (\Gamma_a+\gamma \Gamma_c)(\Gamma_c-\gamma \Gamma_a)
\Big] \\
&-& \frac{4(M-1)\gamma}{(1+M\gamma)^4} \frac{1}{\Gamma_s}
(\Gamma_c+\gamma \Gamma_a)(2\Gamma_c+(M\gamma-1) \Gamma_a) ,
\end{eqnarray*}
where $\Gamma_s = \Gamma_a+\Gamma_c \big( \gamma (M-1) + 1 \big)$.
$\Gamma_{a(c)} = |V_{a(c)}|^2 \pi \rho_{a(c)}$ is the coupling
strength between the impurity and tip (host) conduction electrons,
and $\gamma=\pi^2 |t_c|^2 \rho_a \rho_c$ is a measure of the strength
for the direct tunneling of conduction electrons between the tip and the
host. $\rho_{a(c)}$ is the density of states for noninteracting tip (host)
conduction electrons at the Fermi level.

The current formula in Eqs.~(\ref{currentfinal}) and (\ref{trans})
can be viewed as a generalization of the Landauer-B\"uttiker
formula to the STM case,\cite{Landauer,Buttiker} where $T_{\rm
tr}(\omega)$ is the transmission probability of the electron
tunneling. The first term $T_0$ of the transmission probability
is the direct tunneling between the tip and host, whereas the
rest describe both indirect tunneling of conduction electrons
through the impurity and interference between the two ways of
electron tunneling. For the single-channel case ($M=1$) the
current formula in Eq.~(\ref{currentfinal}) is reduced to the
well-known formula.\cite{Hofstetter} In this case one may expect
$\gamma \ll 1$ due to weakness of the tip coupling, hence $Q_R$
never vanishes. For the multichannel case $\gamma=1/M$ may happen
when the channel number $M$ is large. In this special case $Q_R=0$
and interference contributions to the tunneling current vanish.
When $\gamma=0$, i.e., there is no direct tunneling between the
tip and host, only the last term of the transmission probability
in Eq.~(\ref{trans}) appears, associated with the indirect
tunneling of electrons between the tip and host through the
impurity. This contribution is proportional to the density of
states (DOS) of the impurity. In the Kondo regime the impurity is
screened by conduction electrons of the host, and this many-body
effect must reflect in the DOS of the impurity, hence also in the
tunneling current. In general, the transmission probability
$T_{\rm tr}(\omega)$ is a superposition of the continuous direct
tunneling, indirect tunneling, and their interferences, giving
rise to the Fano resonance.\cite{Fano}

\subsection{Discussion}
\label{subprofile}

The linear conductance is given by
\begin{equation}
G(\omega) = \frac{\partial J(\omega)}{\partial eV}\bigg|_{eV=0} =
\frac{e}{h} \sum_{\sigma\tau} T_{\rm tr}(\omega)
\label{conductance}
\end{equation}
at zero temperature. In general, the presence of the STM tip could
affect physical properties of the host with the Kondo impurity.
We will discuss this point in Subsec.~\ref{subsec}. However, if
couplings of the tip to the host and impurity are weak, influences
of the tip on the host and impurity are negligible, where the
impurity Green function can be evaluated without couplings of the
tip.

\begin{figure}[b]
\includegraphics[width=0.48\textwidth]{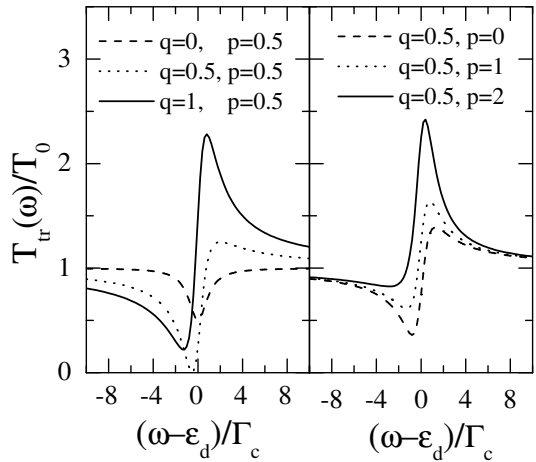}
\caption{Conductance profile in the noninteraction case for
various parameters $q$ and $p$ as indicated in the figure. }
\label{figprofile0}
\end{figure}

The impurity Green function can be written in the form of the
Dyson equation
\begin{equation}
G^{R}_{d\sigma\tau,d\sigma\tau} = \frac{1}{\omega - \varepsilon_d
+ i\Gamma_c - \Sigma(\omega)},
\end{equation}
where $\Sigma(\omega)$ is the retarded self-energy of the impurity
Green function. Considering real and imaginary parts of the
retarded self-energy, $\Sigma(\omega) = \Sigma_R(\omega) - i
\Sigma_I(\omega)$, we obtain the conductance profile in the form
of
\begin{eqnarray}
T_{\rm tr}(\omega) = T_{0} \frac{[\Omega(\omega)+q(\omega)]^2 + p(\omega)}{\Omega^2(\omega)+1} ,
\label{profile}
\end{eqnarray}
where
\begin{eqnarray*}
\Omega(\omega) &=& \frac{\omega-\varepsilon_d-\Sigma_{R}(\omega)}{\Gamma_c+\Sigma_{I}(\omega)} , \\
q(\omega) &=& \frac{(1+M\gamma)^2}{8 \gamma} \frac{Q_R}{\Gamma_c+\Sigma_{I}(\omega)} , \\
p(\omega) &=& 1 - q^2(\omega) - \frac{(1+M\gamma)^2}{4  \gamma} \frac{Q_I}{\Gamma_c+\Sigma_{I}(\omega)}  .
\end{eqnarray*}
In the large frequency limit $\Sigma_R(\omega) \rightarrow {\rm
const.}$ and $\Sigma_I(\omega) \rightarrow 0$ result, thus we have
$T_{\rm tr}(\omega) \rightarrow  T_0$, nothing but the background
profile of the conductance given by the direct tunneling
probability between the tip and host.

The conductance profile in Eq.~(\ref{profile}) can be viewed as a
generalized Fano form. In the noninteraction case of
$\Sigma(\omega)=0$ we have
$\Omega(\omega)=(\omega-\varepsilon_d)/\Gamma_c$, and $p(\omega)$,
$q(\omega)$ are constants. In this case the Fano resonance results
from the interference effect of a Lorentzian line-shape of a
discrete level with a flat continuous background. The quantity $q$
is the so called asymmetry parameter of the Fano line-shape,
whereas the quantity $p$ shifts positions of the maximum and
minimum in the Fano line-shape. In Fig. \ref{figprofile0} we
present the conductance profile for various parameters of $q$ and
$p$. It shows the Fano-resonance line-shape like the Lorentzian
one, the width of which is of order of $\Gamma_c$. For $q=0$ the
profile line-shape is symmetric, and its asymmetry becomes obvious
as $q$ increases. The parameter $p$ not only affects the maximum
and minimum positions of the Fano line-shape, but also makes the
line-shape asymmetry clearer. One can expect that when
interactions are included, the conductance profile will be
significantly modified by the impurity self-energy as well as the
Fano resonance.

\section{Nonequilibrium noncrossing approximation}

The impurity Green function is evaluated within the nonequilibrium
NCA, derived in a similar way as the equilibrium
case.\cite{Parcollet} We start from an Anderson model in the
slave-boson representation
\begin{eqnarray}
\lefteqn{ S_{\rm eff} = \int \int dt dt' \sum_{\sigma\tau}
c^{\dagger}_{\sigma\tau}(t) [
G^{0}_{c\sigma\tau,c\sigma\tau}(t,t')]^{-1}
c_{\sigma\tau}(t') } \nonumber \\
&& + \int dt \sum_{\sigma} f^{\dagger}_{\sigma}(t)
(i\partial_t - \varepsilon_d - i \lambda) f_{\sigma}(t) \nonumber \\
&& +
\int dt \sum_{\tau} b^{\dagger}_{\bar{\tau}}(t)
(i\partial_t - i \lambda) b_{\bar{\tau}}(t)
+ \int dt i \lambda
\nonumber \\
&& + \int dt V_c f^{\dagger}_{\sigma}(t) b_{\bar{\tau}}(t)
c_{\sigma\tau}(t) + {\rm h.c.} ,
\end{eqnarray}
where $G^{0}_{c\sigma\tau,c\sigma\tau}(t,t')$ is the
nonequilibrium single-site noninteracting Green function for host
conduction electrons and time integration is performed along the
Keldysh time contour.


Integrating over conduction electron fields and introducing two
bi-local fields $\Sigma_{f}(t,t')$ and $\Sigma_{b}(t,t')$
conjugate to $\sum_{\sigma} f^{\dagger}_{\sigma}(t)
f_{\sigma}(t')$ and $\sum_{\tau} b^{\dagger}_{\bar{\tau}}(t)
b_{\bar{\tau}}(t')$, respectively, the quartic term in the
effective action can be decoupled as follows
\begin{eqnarray}
\lefteqn{S_{\rm eff} =
\int dt \sum_{\sigma} f^{\dagger}_{\sigma}(t) (i\partial_t - \varepsilon_d - i \lambda) f_{\sigma}(t) } \nonumber \\
&& +
\int dt \sum_{\tau} b^{\dagger}_{\bar{\tau}}(t) (i\partial_t - i \lambda) b_{\bar{\tau}}(t) + \int dt i \lambda \nonumber \\
&&- \int \int dt dt' \Sigma_{f}(t',t) \sum_{\tau}  b^{\dagger}_{\bar{\tau}}(t') b_{\bar{\tau}}(t) \nonumber \\
&& - \int \int dt dt' \Sigma_{b}(t,t') \sum_{\sigma}  f^{\dagger}_{\sigma}(t) f_{\sigma}(t') \nonumber \\
&& - \int \int dt dt' \Sigma_{b}(t,t') D_{0}^{-1}(t,t') \Sigma_{f}(t',t) ,
\end{eqnarray}
where $D_{0}(t,t')=|V_c|^2 G^{0}_{c\sigma\tau,c\sigma\tau}(t,t')$ is the hybridization function.

The nonequilibrium NCA is the saddle-point approximation of the
effective Keldysh action for the bi-local fields
$\Sigma_{f}(t,t')$ and $\Sigma_{b}(t,t')$.
Introducing
nonequilibrium fermionic and bosonic Green functions as $F(t,t')=-
i\langle T_c f_{\sigma}(t) f^{\dagger}_{\sigma}(t') \rangle $ and
$B(t,t')=i \langle T_c b_{\bar{\tau}}(t)
b^{\dagger}_{\bar{\tau}}(t') \rangle $, we find the saddle-point
equations for the fermionic and bosonic self-energies
\begin{eqnarray}
\Sigma_{f}(t,t') &=& i M D_{0}(t,t') B(t,t'), \label{sf} \\
\Sigma_{b}(t,t') &=& - i N D_{0}(t',t) F(t,t') , \label{sb}
\end{eqnarray}
where the bi-local fields play the role of self-energies of the
fermionic and bosonic Green functions in the saddle-point
approximation, given by
\begin{eqnarray}
F^{-1}(t,t') &=& \delta(t,t')(i\partial_t - \varepsilon_d - i\lambda ) - \Sigma_{f}(t,t'), \label{dtf} \\
B^{-1}(t,t') &=& \delta(t,t')(i\partial_t -i\lambda ) - \Sigma_{b}(t,t') . \label{dtb}
\end{eqnarray}

Variation of the effective action with respect to the Lagrange
multiplier $\lambda$ gives rise to the constraint equation
\begin{eqnarray}
N \langle f^{\dagger}_{\sigma} f_{\sigma} \rangle + M \langle b^{\dagger}_{\bar{\tau}} b_{\bar{\tau}} \rangle =1 .
\label{constr}
\end{eqnarray}

Using the Langreth's rule of analytical continuation on the real
time axis,\cite{Langreth,Jauho} the self-energy equations
(\ref{sf})-(\ref{sb}) are
\begin{eqnarray}
\Sigma_{f}^{R}(t,t') &=& i M \big[
\big( D_{0}^{R}(t,t')+ D_{0}^{<}(t,t') \big) B^{R}(t,t') \nonumber \\
&& + D_{0}^{R}(t,t') B^{<}(t,t')
 \big] , \label{srf} \\
 \Sigma_{b}^{R}(t,t') &=& - i N \big[
D_{0}^{<}(t',t) F^{R}(t,t') \nonumber \\
&& + D_{0}^{A}(t',t) F^{<}(t,t')
 \big] , \label{srb} \\
  \Sigma_{f}^{<}(t,t') &=& i M D_{0}^{<}(t,t') B^{<}(t,t') , \label{slf} \\
 \Sigma_{b}^{<}(t,t') &=& - i N D_{0}^{>}(t',t) F^{<}(t,t') . \label{slb}
\end{eqnarray}
In the steady state the Fourier transformation for the Green
functions and their self-energies results in the following
nonequilibrium NCA equations
\begin{eqnarray}
\Sigma_{f}^{R}(\omega) &=& M \Gamma_c \int \frac{d\varepsilon}{2\pi} B^{<}(\varepsilon) \nonumber \\
&& + M \Gamma_c \int \frac{d\varepsilon}{\pi} f_c(\varepsilon-\omega) B^{R}(\varepsilon) , \label{srfw} \\
\Sigma_{b}^{R}(\omega) &=& N \Gamma_c \int \frac{d\varepsilon}{2\pi} F^{<}(\varepsilon) \nonumber \\
&& + N \Gamma_c \int \frac{d\varepsilon}{\pi} f_c(\varepsilon-\omega) F^{R}(\varepsilon) , \label{srbw} \\
\Sigma_{f}^{<}(\omega) &=& - M \int \frac{d\varepsilon}{\pi} f_c(\omega-\varepsilon) B^{<}(\varepsilon), \label{slfw} \\
\Sigma_{b}^{<}(\omega) &=& - N \int \frac{d\varepsilon}{\pi}
f_c(\omega-\varepsilon) F^{<}(\varepsilon) , \label{slbw}
\end{eqnarray}
where we have used explicit expressions for the hybridization
function
\begin{eqnarray}
D_{0}^{R}(\omega) &=& |V_c|^2 G^{0 R}_{c\sigma\tau,c\sigma\tau}(\omega) = - i \Gamma_c , \\
D_{0}^{<}(\omega) &=& |V_c|^2 G^{0
<}_{c\sigma\tau,c\sigma\tau}(\omega) = 2 i \Gamma_c f_c(\omega) .
\end{eqnarray}
Note that the first terms in Eqs.~(\ref{srfw}) and (\ref{srbw}) are
just constants. They can be absorbed into the Lagrange multiplier,
using the constraint equation (\ref{constr}) and
\begin{eqnarray}
\langle f^{\dagger}_{\sigma} f_{\sigma} \rangle &=& -i \int \frac{d\varepsilon}{2\pi} F^{<}(\varepsilon) , \\
\langle b^{\dagger}_{\bar{\tau}} b_{\bar{\tau}} \rangle &=& i \int
\frac{d\varepsilon}{2\pi} B^{<}(\varepsilon) .
\end{eqnarray}
We also used the fact that the Lagrange multiplier takes a large
value at the end of calculations.\cite{Wingreen,Hettler}

The Dyson equations (\ref{dtf})-(\ref{dtb}) can be also rewritten
for the retarded and lesser Green functions based on the
Langreth's rule of analytical continuation
\begin{eqnarray}
F^{R}(\omega) &=& \frac{1}{\omega-\varepsilon_d - i \lambda - \Sigma^{R}_{f}(\omega) } , \label{d1} \\
B^{R}(\omega) &=& \frac{1}{\omega - i \lambda - \Sigma^{R}_{b}(\omega) } , \label{d2} \\
F^{<}(\omega) &=&  F^{R}(\omega) \Sigma^{<}_{f}(\omega) F^{A}(\omega) , \label{d3} \\
B^{<}(\omega) &=&  B^{R}(\omega) \Sigma^{<}_{b}(\omega)
B^{A}(\omega) . \label{d4}
\end{eqnarray}

Finally, the impurity Green function can be calculated via the fermionic and bosonic
Green functions
\begin{equation}
G^{R}_{d\sigma\tau,d\sigma\tau}(\omega) =
i \int \frac{d\varepsilon}{2\pi} \big[
F^{<}(\omega+\varepsilon) B^{A}(\varepsilon) + F^{R}(\omega+\varepsilon) B^{<}(\varepsilon)
\big] .
\label{gds}
\end{equation}
Inserting this impurity Green function into the transmission
coefficient, we find the conductance profile measured in STM.

The present derivation of nonequilibrium NCA equations can be
viewed as the path integral version for the projection
method,\cite{Wingreen,Hettler} completely equivalent with each
other. In practice, such NCA equations are first solved for
retarded Green functions, and lesser Green functions are found
with the use of the retarded Green functions. In the next section
we will perform this work.

\section{Fano resonance in the multi-channel Kondo effect}

\subsection{Zero temperature solution of the noncrossing approximation equations}

In the linear response regime the host-electron distribution
function $f_c(\omega)$ is given by the standard Fermi-Dirac
distribution function. Then, the NCA equations
(\ref{srfw})-(\ref{srbw}) for the retarded Green functions
resemble the equilibrium NCA equations, solved exactly at zero
temperature.\cite{Muller,Bickers} Eqs.~(\ref{srfw})-(\ref{srbw})
can be written as the following differential equations at zero
temperature
\begin{eqnarray}
\frac{d \Sigma^{R}_{f}(\omega)}{d\omega} &=& \frac{M \Gamma_c}{\pi} B^{R}(\omega), \label{desf}\\
\frac{d \Sigma^{R}_{b}(\omega)}{d\omega} &=& \frac{N \Gamma_c}{\pi} F^{R}(\omega) , \label{desb}
\end{eqnarray}
with the boundary condition
$\Sigma^{R}_{f}(-D)=\Sigma^{R}_{b}(-D)=0$, where $D$ is the band
cutoff.

Solving these NCA equations, one can find \begin{equation}
-\big[F^R(\omega)\big]^{-1} = T_K h_{\frac{N}{M}}
\Big[\frac{E_0-\omega}{T_K}, \frac{T_K}{\Gamma_c}  \Big] ,
\label{yfs}
\end{equation}
where the scaling function $h_{\alpha}[x]$ is given by
\begin{eqnarray}
x = \int_{0}^{h_{\alpha}[x,c]} d y \frac{W[y^{\alpha} e^{\pi
\alpha c y}]}{1 + W[y^{\alpha} e^{\pi \alpha c y}]}.
\label{scaling}
\end{eqnarray}
$W[x]$ is the Lambert $W$ function, defined as\cite{Corless}
$$
x=W[x] \exp(W[x]) .
$$
$T_K = D [ M \Gamma_c / \pi D ]^{M/N} \exp[\pi \varepsilon_d/N
\Gamma_c]$ is identified with the Kondo energy scale, below which
the multichannel Kondo effect arises. $E_{0}$ the ground state
energy of the impurity, below which spectral densities of the
fermionic and bosonic fields vanish at zero
temperature.\cite{Muller,Bickers} Therefore, $F^R(\omega)$,
$B^{R}(\omega)$ are real functions below $E_0$. As shown in this
expression, there are two energy scales $\Gamma_c$ and $T_K$ for
the NCA solution. Detailed derivation can be found in Appendix B.

Equation~(\ref{yfs}) shows that the NCA solution obeys the
universal scaling form. Although the scaling function in
Eq.~(\ref{scaling}) should be computed numerically, its asymptotes
in limits $x\ll 1$ and $x \gg 1$ can be found analytically. For
$x\ll 1$, $W[x] = x - x^2$, thus we obtain the asymptote
\begin{eqnarray}
h_{\alpha}[x,c] &=& [(\alpha+1) x]^{1/(\alpha+1)} \Big[ 1 -
\frac{\pi \alpha c}{\alpha + 2} [(\alpha+1) x]^{1/(\alpha+1)} \nonumber \\
&& + \frac{2}{2 \alpha+1} [(\alpha+1) x]^{\alpha/(\alpha+1)}
\Big] . \label{asympt}
\end{eqnarray}
The leading term of the scaling function $h_{\alpha}[x,c]$ shows
the power scaling with an exponent $1/(\alpha+1)$, implying that
both the retarded fermionic and bosonic Green functions exhibit
power-law physics near the threshold energy $E_0$. This
corresponds to the overcompensated regime, where the impurity spin
is over-screened by multichannel conduction
electrons.\cite{Affleck,Cox,Parcollet} In the opposite limit $x
\gg 1$, we obtain $h_{\alpha}[x,c] = x$, leading the fermionic
Green function behave like $1/\omega$. This corresponds to the
free moment regime, where the impurity spin is weakly bound to
screening clouds.

\begin{figure}[t]
\includegraphics[width=0.48\textwidth]{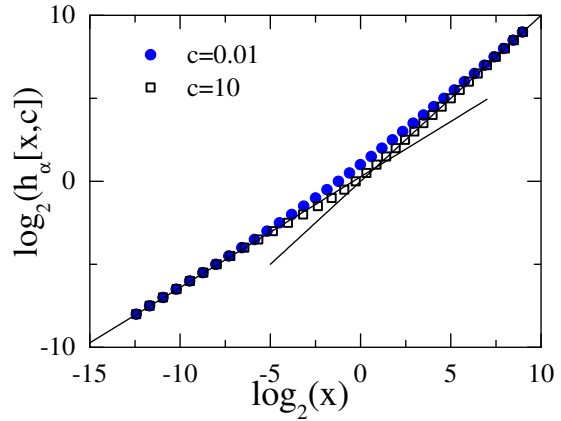}
\caption{(Color online) Scaling function $h_{\alpha}[x,c]$ with
$\alpha=0.5$ and various parameters $c$. The two asymptotes
$[(\alpha+1) x]^{1/(\alpha+1)}$ and $x$ are shown as the solid
lines.} \label{figsc}
\end{figure}

In Fig.~\ref{figsc} we plot the scaling function
$h_{\alpha}[x,c]$. This NCA solution
[Eqs.~(\ref{bviaf})-(\ref{yfs})] may be viewed as the complete
solution of the NCA equations at zero temperature for all energy
scales below the high-energy cutoff. It shows that the scaling
function $h_{\alpha}[x,c]$ crosses from the power scaling regime
to the linear behavior as $x$ increases from zero. The scaling
function $h_{\alpha}[x,c^*]$ obeys the power law up to $x
\approx 1$, where $c^{*}$ is identified with a "crossover" value.
On the other hand, the power law of the scaling function is valid
only for $x\ll 1$ when $c \ll c^{*}$. This property implies that
there is a characteristic value of $(T_K/\Gamma_c)^{*}$, leading
the power scaling to persist until energies comparable to $T_K$.
In the conventional case of $T_K/\Gamma_c \ll 1$ the power scaling
holds only for energies much below $T_K$. In the scaling regime
$F^{R}(\omega) \sim (|\omega-E_0|/T_K)^{M/(N+M)}$ and
$B^{R}(\omega) \sim (|\omega-E_0|/T_K)^{N/(N+M)}$ result, obtained
previously.\cite{Muller,Bickers} Although such power-law scaling
fails to describe Fermi liquid in $T < T_{K}$ for the
single-channel case, it is the underlying physics of the
overcompensated screening impurity in the multichannel case, as
shown by the conformal field theory.\cite{Affleck}

The lesser self-energies obey the following differential equations
\begin{eqnarray}
\frac{d \Sigma^{<}_{f}(\omega)}{d\omega} &=& \frac{M \Gamma_c}{\pi} B^{<}(\omega), \label{les1} \\
\frac{d \Sigma^{<}_{b}(\omega)}{d\omega} &=& \frac{N
\Gamma_c}{\pi} F^{<}(\omega)   \label{les2}
\end{eqnarray}
at zero temperature. In the scaling regime the lesser Green
functions also display the same power scaling as the retarded
ones, given by\cite{Muller,Bickers}
\begin{eqnarray}
F^{<}(\omega) &=& i A \frac{1}{Y_f(\omega)} , \label{ls1} \\
B^{<}(\omega) &=& - i A \frac{1}{Y_b(\omega)} \label{ls2}
\end{eqnarray}
with $A=2\pi/(N+M)$. See Appendix B. Based on Eq.~(\ref{yfs}) with
Eq.~(\ref{bviaf}) and Eqs.~(\ref{ls1}), (\ref{ls2}), we find the
final expression for the impurity Green function from
Eq.~(\ref{gds}) in the scaling regime
\begin{eqnarray}
\lefteqn{
G^{R}_{d\sigma\tau,d\sigma\tau}(\omega) = G^{R}_{d\sigma\tau,d\sigma\tau}(0) + \Delta G_{d\sigma\tau,d\sigma\tau}(\omega) , }
\label{gimp}\\
&& \hspace{-0.4cm} G^{R}_{d\sigma\tau,d\sigma\tau}(0) = \frac{1}{N+M} \frac{\pi}{\Gamma_c}
 \Big[ \frac{1}{M} - \frac{N+M}{N} n_f - \nonumber \\
&& \frac{\pi}{N+M} \cot \Big(\frac{\pi M}{N+M}\Big) - i \frac{\pi}{N+M} \Big], \nonumber \\
&& \hspace{-0.4cm} \Delta G_{d\sigma\tau,d\sigma\tau}(\omega) = i \frac{4 \pi}{(N+M)\Gamma_c} \sin\Big(\frac{\pi M}{N+M}\Big)
\nonumber \\
&& \hspace{-0.4cm} \Big[\frac{M}{2N+M}
 B\Big(\frac{2N}{N+M},\frac{M}{N+M}\Big) \Big(- \frac{N+M}{M} \frac{\omega}{T_K} \Big)^{\frac{N}{N+M}}  \nonumber \\
&& \hspace{-0.4cm} + \frac{\pi T_K}{(N+2 M) \Gamma_c}
 B\Big(\frac{N}{N+M},\frac{2 M}{N+M}\Big) \Big( \frac{N+M}{M} \frac{\omega}{T_K} \Big)^{\frac{M}{N+M}}
\Big] , \nonumber
\end{eqnarray}
where $B(x,y)$ is the beta function,\cite{Stegun} and $n_f=\langle
f^{\dagger}_{\sigma} f_{\sigma} \rangle$.

The impurity Green function exhibits two power scalings with
$N/(M+N)$ and $M/(N+M)$. In the overcompensation case ($M>N$) the
power $N/(N+M)$ scaling is dominant for $\omega < T_{NCA}$, where
$T_{NCA}$ is the crossover energy when the dominant scaling
behavior of the impurity Green function crosses from one power to
another, given by
\begin{eqnarray}
T_{NCA} &=& T_K \frac{M}{N+M} \Bigg[
\frac{M(N+2M)\Gamma_c}{(2N+M)\pi T_K} \frac{B\Big(\frac{2N}{N+M},\frac{M}{N+M}\Big)}{B\Big(\frac{N}{N+M},\frac{2 M}{N+M}\Big)}
\nonumber \\
&& \times \cos\Big( \frac{\pi M}{N+M} \Big)
\Bigg]^{\frac{N+M}{M-N}} .
\end{eqnarray}

\subsection{Conductance profile within the noncrossing approximation}

The impurity Green function Eq.~(\ref{gimp}) is nonanalytic at
$\omega=0$, exhibiting an asymmetric and sharp cusp with power-law
scaling around $\omega=0$. These non-Fermi liquid features are
reflected on the conductance profile in Eq.~(\ref{profile}). For
comparison, we also calculate the conductance profile in the
Fermi-liquid phase, given by the following
self-energy\cite{Yosida}
\begin{eqnarray}
\Sigma_{\rm FL}(\omega) = - \Gamma_{c} \Big[ \frac{\omega}{T_{K}} + i \frac{1}{2} \Big( \frac{\omega}{T_K} \Big)^2 \Big] .
\label{fermi}
\end{eqnarray}

\begin{figure}[t]
\includegraphics[width=0.48\textwidth]{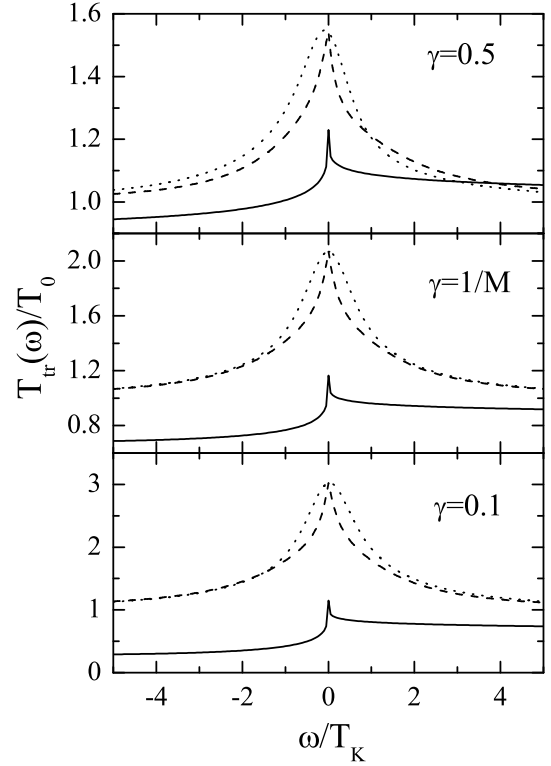}
\caption{Conductance profile calculated within the NCA (the solid
lines), the marginal Fermi-liquid theory (the dashed lines), and
the Fermi-liquid theory (the doted lines) for various values of
$\gamma$. Other parameters are $M=6$, $N=2$, $\Gamma_c=1$,
$\Gamma_a=0.01$, $T_K=0.01$, and $n_f=0.8$.} \label{fignca}
\end{figure}

In Fig. \ref{fignca} we plot conductance profiles in the symmetric
case, i.e., $\varepsilon_d=-{\rm Re} \Sigma(0)$, within the NCA
and Fermi-liquid theory for various values of $\gamma$. It shows
that the conductance profile of the overcompensation multichannel
Kondo model shows a sharp cusp with power-law scaling at
$\omega=0$, as expected. This feature is completely distinguished
from the Fermi-liquid theory result, where the conductance profile
exhibits the narrow Fano-Kondo resonance, the width of which is of
order of $T_K$. Note that the frequency in the $x$-axis is scaled
with $T_{K}$ in Fig.~\ref{fignca}, which is much smaller than the
energy scale $\Gamma_c$ in the noninteraction case (Fig.
\ref{figprofile0}).

In the case of $\gamma=1/M$ the asymmetry parameter $q(\omega)$
vanishes, thus interference contributions to the conductance
disappear. As we can see in Fig.~\ref{fignca}, the Fermi-liquid
conductance profile is symmetric and exhibits only the Kondo
resonance at zero frequency. On the other hand, the asymmetry
parameter $q(\omega)$ is finite for $\gamma \neq 1/M$, and the
Fano resonance shifts the peak position away from zero, giving
rise to an asymmetric feature in the conductance profile within
the Fermi-liquid theory. Note that the asymmetry is more
pronounced for larger $\gamma$.

In the multichannel case the conductance profile still exhibits an
asymmetric feature even when $\gamma=1/M$. However, this asymmetry
is not due to the Fano resonance, but due to the non-Fermi liquid
feature in the density of states of the overscreened impurity.
Although interference contributions to the conductance are finite
for $\gamma \neq 1/M$, they cannot shift the peak position away
from zero in the multichannel conductance as in the Fermi liquid
theory, the hallmark of non-Fermi liquid physics measured in STM.
The conductance profile still exhibits the sharp cusp at
$\omega=0$. This asymmetry is due to both interference
contributions and non-Fermi liquid properties of the overscreened
impurity. However, interference contributions are so weak that the
Fano resonance is suppressed. This indicates dominance of the
non-Fermi liquid overcompensation over the Fano resonance. In the
multichannel case the conductance profile shows weak dependence on
$\gamma$, thus the strength of the tip coupling is not essential
to detect the non-Fermi liquid feature in the tunneling
conductance, as far as it does not vanish.

\subsection{Physical origin for suppression of the Fano resonance}
\label{subsec}

In the Fermi liquid phase one can see $\Omega(\omega) \sim \omega
$ and $q(\omega)$, $p(\omega) \sim {\rm const}$ as $\omega
\rightarrow 0$ in the tunneling conductance [Eq. (\ref{profile})]
from Eq.~(\ref{fermi}), giving rise to the Fano resonance away
from $\omega=0$. This originates from quantum coherence of
impurity dynamics, which maintains $\Sigma_{R}(\omega) \sim
\omega$ and $\Sigma_{I}(\omega) \sim \omega^2$ as $\omega
\rightarrow 0$. In the multichannel overcompensation scaling
regime $\Sigma_{R}(\omega) \sim \Sigma_{I}(\omega) \sim
|\omega|^{N/(N+M)}$ results when $\omega \rightarrow 0$, which
leads $\Omega(\omega) \sim 1$ and the asymmetry factor $q(\omega)
\sim |\omega|^{-N/(N+M)}$, $p(\omega) \sim |\omega|^{-N/(N+M)}$.
As a consequence we find $T_{\rm tr}(\omega) \sim
|\omega|^{-2N/(N+M)}$ in $\omega \rightarrow 0$. The conductance
profile always exhibits its peak at $\omega=0$ that indicates the
peak position does not shift unlike the Fermi-liquid theory. This
can be interpreted as the fact that quantum coherence of impurity
dynamics is lost in the multichannel scaling regime, hence
suppressing the Fano resonance.

For comparison we also calculate the conductance profile in a
marginal Fermi-liquid phase, given by the following self-energy
ansatz\cite{Varma}
\begin{eqnarray}
\Sigma_{\rm MFL}(\omega) =  \Gamma_{c} \Big[ \frac{\omega}{T_{K}} \log \frac{|\omega|}{T_{K}}
 - i \frac{\pi}{2} \frac{|\omega|}{T_K}  \Big] .
\end{eqnarray}
The marginal Fermi-liquid phase can be considered as an
intermediate case which lies between the Fermi-liquid phase and
multichannel overcompensation phase, where the impurity dynamics
maintains weak quantum coherence. In Fig.~\ref{fignca} we also
plot the conductance profile in the marginal Fermi-liquid phase
for various values of $\gamma$. It shows that the conductance
profile of the marginal Fermi-liquid theory exhibits a narrow
resonance nearby $\omega=0$. It resembles to the conductance
profile of the Fermi-liquid theory, except for the position of the
Fano resonance which very closes to $\omega=0$. The weak quantum
coherence in the marginal Fermi-liquid phase can maintain the Fano
resonance, however, weakness of its quantum coherence cannot
significantly shift the peak position of the Fano resonance away
from zero in comparison with the Fermi-liquid phase. This feature
implies the important role of quantum coherence of impurity
dynamics in the mechanism of Fano resonance. When the impurity
dynamics becomes incoherent, the Fano resonance is suppressed.

\subsection{Effect of the tip-host and the tip-impurity coupling on the
$M$-channel Kondo impurity system}

As mentioned in Subsec.~\ref{subprofile}, the impurity Green
function is evaluated without the tip coupling in the conductance
profile of Eq.~(\ref{profile}). In general, the presence of the
tip could affect physical properties of the host with the Kondo
impurity. In this subsection we discuss the effects of the
tip-host and the tip-impurity coupling on the $M$-channel Kondo
impurity system of the host and impurity. Although such tip
couplings break the $SU(M)$ symmetry in principle, we argue that
the overcompensation Kondo effect still occurs for weak
tip-impurity couplings with finite tip-host couplings.

\subsubsection{Effect of the tip-host coupling}
\label{subsub1} First, we consider the effect of the tip-host
coupling only, where the tip-impurity coupling is neglected. We
take the following unitary transformation for conduction electron
fields $c_{\mathbf{k}\sigma\tau}$ and bosonic fields $b_{\tau}$
\begin{eqnarray}
c_{\mathbf{k}\sigma\tau} &=& \sum_{\tau'} \mathbf{U}_{\tau\tau'}
\tilde{c}_{\mathbf{k}\sigma\tau'} , \label{uc}\\
b_{\tau} &=& \sum_{\tau'} \mathbf{U}_{\tau\tau'}^{\dagger}
\tilde{b}_{\tau'}, \label{ub}
\end{eqnarray}
where the $M\times M$ unitary matrix $\mathbf{U}$ is chosen to
satisfy $\tilde{c}_{\mathbf{k}\sigma
1}=\sum_{\tau}c_{\mathbf{k}\sigma\tau}/\sqrt{M}$ for
diagonalization of the tip to host coupling term. We rewrite the
starting Hamiltonian Eq.~(\ref{ham}) in the above transformed
basis and integrate over tip conduction-electron fields. Then, we
find
\begin{eqnarray*}
\lefteqn{ S_{\rm eff} = \int d\tau \sum_{\mathbf{k}\sigma\tau}
\tilde{c}^{\dagger}_{\mathbf{k}\sigma\tau}(\tau)
( \partial_{\tau} - \varepsilon_{\mathbf{k}} ) \tilde{c}_{\mathbf{k}\sigma\tau}(\tau) }  \\
&&+ \sum_{\sigma} f^{\dagger}_{\sigma}(\tau) ( \partial_{\tau} +
\lambda - \varepsilon_{f} ) f_{\sigma}(\tau) + \sum_{\tau}
\tilde{b}^{\dagger}_{\bar{\tau}}(\tau) ( \partial_{\tau} + \lambda
) \tilde{b}_{\bar{\tau}}(\tau) \\
&&+ V_c \sum_{\mathbf{k}\sigma\tau} f^{\dagger}_{\sigma}(\tau)
\tilde{b}_{\bar{\tau}}(\tau)
\tilde{c}_{\mathbf{k}\sigma\tau}(\tau)
+ {\rm h.c.} \\
&&+ \int d\tau d\tau' M |t_c|^2 \sum_{\mathbf{k} \mathbf{k}'
\sigma} \tilde{c}^{\dagger}_{\mathbf{k}\sigma 1}(\tau)
g_{a}(\tau-\tau') \tilde{c}_{\mathbf{k}'\sigma 1}(\tau') ,
\end{eqnarray*}
where $g_{a}(\tau-\tau')$ is the local Green function for tip
electrons.

It is clear that the last term in the above effective action
breaks the $SU(M)$ symmetry. Effectively, the conduction channel
$\tilde{c}_{\mathbf{k}\sigma 1}$ has an additional contribution
$\sim M |t_c|^2 \rho_a$ to its normal dispersion, while other
conduction channels have not. Integrating over conduction
electrons, we obtain
\begin{eqnarray}
\lefteqn{ S_{\rm eff} = \int d\tau \sum_{\sigma}
f^{\dagger}_{\sigma}(\tau) ( \partial_{\tau} + \lambda -
\varepsilon_{f} ) f_{\sigma}(\tau) } \\ &&+ \sum_{\tau}
\tilde{b}^{\dagger}_{\bar{\tau}}(\tau) (
\partial_{\tau} + \lambda
) \tilde{b}_{\bar{\tau}}(\tau)  + \int d\tau d\tau' |V_{c}|^{2} \nonumber \\
&& \sum_{\sigma} f^{\dagger}_{\sigma}(\tau) \tilde{b}_{1}(\tau)
\Bigl\{ \sum_{\mathbf{k} \mathbf{k}'} g^{c (1)}_{kk'}(\tau-\tau')
\Bigr\} \tilde{b}_{1}^{\dagger}(\tau') f_{\sigma}(\tau') \nonumber \\
&&+ \int d\tau d\tau' |V_{c}|^{2}  \sum_{\sigma,\tau\not= 1}
f^{\dagger}_{\sigma}(\tau) \tilde{b}_{\tau}(\tau) \Bigl\{
\sum_{\mathbf{k} \mathbf{k}'} g^{c}_{kk'}(\tau-\tau') \Bigr\} \nonumber \\
&& \hspace{3cm} \tilde{b}_{\tau}^{\dagger}(\tau')
f_{\sigma}(\tau') , \label{effaction}
\end{eqnarray}
where $[g^{c (1)}_{kk'}(\tau-\tau')]^{-1} = - ( \partial_{\tau} -
\varepsilon_{\mathbf{k}} )\delta_{kk'} -  M |t_c|^2
g_{a}(\tau-\tau')$ is the inverse of the electron propagator for
the channel $\tau = 1$ and $[g^{c}_{kk'}(\tau-\tau')]^{-1} = - (
\partial_{\tau} - \varepsilon_{\mathbf{k}} )\delta_{kk'}$ is that
for other $M-1$ channels. This $SU(M)$ symmetry breaking gives
rise to anisotropic hybridization couplings. In Appendix C we
prove that $\Gamma^{(1)}_{c} = |V_c|^2
\sum_{\mathbf{k}\mathbf{k}'}|{\rm Im}g^{c(1)}_{kk'}(0)| \leq
\Gamma_c$ always happens. This indicates that the channel $\tau=1$
couples to the impurity weaker than $(M-1)$ rest channels. As a
consequence, the $(M-1)$ channel Kondo effect would occur. For
large $M$, there is no difference in physics of the
overcompensated Kondo effect between $(M-1)$ channels and $M$
channels.

\subsubsection{Effect of the tip-impurity coupling}
\label{subsub2} Now we consider the effect of the tip-impurity
coupling only, where the tip-host coupling is neglected. As
performed in the previous subsubsection, we take again the unitary
transformation in Eqs.~(\ref{uc})-(\ref{ub}) for conduction
electron fields $c_{\mathbf{k}\sigma\tau}$ and bosonic fields
$b_{\tau}$, but now the $M\times M$ unitary matrix $\mathbf{U}$ is
chosen to satisfy $\tilde{b}_{1}=\sum_{\tau}b_{\tau}/\sqrt{M}$ for
diagonalization of the tip to impurity coupling term. Then, we
find the following effective action
\begin{eqnarray*}
\lefteqn{ S_{\rm eff} = \int d\tau  \sum_{\sigma}
f^{\dagger}_{\sigma}(\tau) ( \partial_{\tau} + \lambda -
\varepsilon_{f} ) f_{\sigma}(\tau) } \nonumber \\
&& + \sum_{\tau} \tilde{b}^{\dagger}_{\bar{\tau}}(\tau) (
\partial_{\tau} + \lambda
) \tilde{b}_{\bar{\tau}}(\tau) + \int d\tau d\tau' \sum_{\sigma}
f^{\dagger}_{\sigma}(\tau) \tilde{b}_{1}(\tau) \nonumber \\
&& \big(|V_c|^2 g_{c}(\tau-\tau')+ M |V_a|^2 g_{a}(\tau-\tau')
\big) \tilde{b}_{1}^{\dagger}(\tau')
f_{\sigma}(\tau') \nonumber \\
&&+ \int d\tau d\tau' |V_{c}|^{2}  \sum_{\sigma,\tau\not= 1}
f^{\dagger}_{\sigma}(\tau) \tilde{b}_{\tau}(\tau)
g_{c}(\tau-\tau') \tilde{b}_{\tau}^{\dagger}(\tau')
f_{\sigma}(\tau') ,
\end{eqnarray*}
where
$g_{c}(\tau)=\sum_{\mathbf{k},\mathbf{k}'}g^{c}_{kk'}(\tau)$. This
effective action shows that the tip-impurity coupling breaks the
$SU(M)$ symmetry. The effective hybridization coupling of the
effective channel $\tau=1$ is $\Gamma_c + M \Gamma_a$, and it is
always larger than the effective hybridization coupling $\Gamma_c$
of the $(M-1)$ rest channels. As a consequence, the single-channel
Kondo effect would occur. In this case the system restores the
Fermi-liquid behaviors. The NCA at zero temperature produces
spurious non-Fermi liquid features,\cite{Muller,Bickers} and it is
not adequate to describe the physical properties of the system. In
the calculation of the conductance profile the Fermi-liquid self
energy in Eq.~(\ref{fermi}) must be used, and the Fano-Kondo
resonance could appear.

\subsubsection{Effect of both the tip-host and tip-impurity couplings}
\label{subsub3}
The two previous subsubsections show that the
tip-host coupling allows the overcompensation Kondo effect, while
the tip-impurity coupling drives the system away from the critical
regime. Therefore, we are mainly interested in the case of a
finite tip-host coupling and small tip-impurity coupling. As the
case of finite tip-host couplings, we take the unitary
transformation for conduction electron fields
$c_{\mathbf{k}\sigma\tau}$ and bosonic fields $b_{\tau}$ in
Eqs.~(\ref{uc})-(\ref{ub}) with $\tilde{c}_{\mathbf{k}\sigma
1}=\sum_{\tau}c_{\mathbf{k}\sigma\tau}/\sqrt{M}$. Proceeding in a
similar way as the previous subsubsections, we find the effective
action
\begin{eqnarray}
\lefteqn{ S_{\rm eff} = \int d\tau \sum_{\sigma}
f^{\dagger}_{\sigma}(\tau) ( \partial_{\tau} + \lambda -
\varepsilon_{f} ) f_{\sigma}(\tau) } \nonumber \\
&&+ \sum_{\tau} \tilde{b}^{\dagger}_{\bar{\tau}}(\tau) (
\partial_{\tau} + \lambda
) \tilde{b}_{\bar{\tau}}(\tau)  +  \sum_{\sigma,\tau\not= 1} \int
d\tau d\tau'  |V_{c}|^{2} \nonumber \\
 && f^{\dagger}_{\sigma}(\tau)
\tilde{b}_{\tau}(\tau)
 g_{c}(\tau-\tau')
\tilde{b}_{\tau}^{\dagger}(\tau') f_{\sigma}(\tau') +
\sum_{\sigma} \int d\tau d\tau' \nonumber \\
&& \Bigl\{ |V_{c}|^{2} f^{\dagger}_{\sigma}(\tau)
\tilde{b}_{1}(\tau) g^{c(1)}(\tau-\tau')
\tilde{b}_{1}^{\dagger}(\tau') f_{\sigma}(\tau') +
\nonumber \\
&& f^{\dagger}_{\sigma}(\tau) b_{0}(\tau) \big[ M |V_a|^2
g_{a}(\tau-\tau') + M^2 |V_a|^2 |t_c|^2 \int d\tau_{1} d\tau_{2}
\nonumber \\
&&  g_{a}(\tau-\tau_{1}) g^{c(1)}(\tau_{1}-\tau_{2})
g_{a}(\tau_{2}-\tau') \big] b_{0}^{\dagger}(\tau')
f_{\sigma}(\tau') +
\nonumber \\
&& M V_c V_a^{*} t_c f^{\dagger}_{\sigma}(\tau)
\tilde{b}_{1}(\tau) \int d\tau_1 g^{c(1)}(\tau-\tau_{1})
g_{a}(\tau_{1}-\tau') \nonumber \\
&& b_{0}^{\dagger}(\tau') f_{\sigma}(\tau') + M V_c^{*} V_a
t_c^{*} f^{\dagger}_{\sigma}(\tau) b_{0}(\tau) \int d\tau_1
g_{a}(\tau-\tau_{1}) \nonumber \\
&& g^{c(1)}(\tau_{1}-\tau')\tilde{b}_{1}^{\dagger}(\tau')
f_{\sigma}(\tau') \Bigr\}
  , \label{effaction3}
\end{eqnarray}
where
$g^{c(1)}(\tau)=\sum_{\mathbf{k},\mathbf{k}'}g^{c(1)}_{kk'}(\tau)$,
and $b_{0}=\sum_{\tau} b_{\tau} / \sqrt{M}$. In general, $b_{0}$
is not necessarily equal to $\tilde{b}_{1}$. We can rewrite
$b_{0}=\alpha_{1} \tilde{b}_1 + \alpha_{2} \hat{b}$, where
$\hat{b}$ is a linear combination of $\tilde{b}_2, \ldots,
\tilde{b}_{M}$. The coefficients $\alpha_1$ and $\alpha_2$ satisfy
$|\alpha_{1}|^2+|\alpha_{2}|^2 =1$.

In the wide band limit of a constant density of states for
conduction electrons in both the tip and host, the effective
hybridization coupling terms of $M$ channels in
Eq.~(\ref{effaction3}) can be rewritten schematically as follows
\begin{eqnarray}
\lefteqn{f^{\dagger}_{\sigma}\tilde{b}_{1} \big( \Gamma_{c}^{(1)}
+ |\alpha_{1}|^2 \Gamma_{a}^{(0)} + \alpha_{1}^{*} \Gamma_{ac}  +
\alpha_{1} \Gamma_{ac}^{*} \big) \tilde{b}_{1}^{\dagger}
f_{\sigma} } \nonumber \\
&& + f^{\dagger}_{\sigma}\hat{b} |\alpha_{2}|^2 \Gamma_{a}^{(0)}
\hat{b}^{\dagger} f_{\sigma} + f^{\dagger}_{\sigma}\tilde{b}_{1}
\alpha_{2}^{*} \Gamma_{ac}
\hat{b}^{\dagger} f_{\sigma}  \nonumber \\
&& + f^{\dagger}_{\sigma}\hat{b} \alpha_{2} \Gamma_{ac}^{*}
\tilde{b}_{1}^{\dagger} f_{\sigma}+ \sum_{\tau\not=1}
f^{\dagger}_{\sigma}\tilde{b}_{\tau} \Gamma_{c}
\tilde{b}_{\tau}^{\dagger} f_{\sigma}
  , \label{effaction3a}
\end{eqnarray}
where
\begin{eqnarray*}
\Gamma_{a}^{(0)} &=& M |V_a|^2 |{\rm Im} g_{a}(0)| + M^2 |V_a|^2
|t_c|^2 |{\rm Im}[ g_{a}(0)^{2} g^{c(1)}(0)]| , \\
 \Gamma_{ac} &=&
M V_c V_a^{*} t_c |{\rm Im}[g_{a}(0) g^{c(1)}(0)]|.
\end{eqnarray*}
One can show that $\Gamma^{(1)}_{c}=\Gamma_c/(1+M \gamma)$,
$\Gamma_{a}^{(0)}= M \Gamma_{a}/(1+M\gamma)$, and $|\Gamma_{ac}|^2
= M^2 \Gamma_c \Gamma_a \gamma /(1+M \gamma)^2$. The hybridization
coupling of $(M-1)$ channels $\Gamma_c$ is expected to dominate
over those of the channels $\tilde{b}_1$ and $\hat{b}$ if the
following conditions are satisfied
\begin{eqnarray}
\Gamma_c &>& \Gamma_{c}^{(1)} + \Gamma_{a}^{(0)} + 2 |\Gamma_{ac}|
,
\\
\Gamma_c^{(1)} &>& \Gamma_{a}^{(0)} .
\end{eqnarray}
These conditions are equivalent to $\Gamma_c > M \Gamma_{a}$ and
$\gamma \Gamma_c > (3+\sqrt{8}) \Gamma_a$. When the hybridization
couplings of $(M-1)$ channels are relevant, the $(M-1)$ channel
Kondo effect would occur. For large $M$, there is again no
difference in physics of the overcompensated Kondo effect between
$(M-1)$ channels and $M$ channels. However, we admit that this
expectation does not have a firm ground because there exists
complex mixing between channels. In our opinion this problem
should deserve further investigation.

Finally, we would like to mention that our demonstration is
expected to apply to the quantum dot realization for the
multi-channel Kondo effect. If the STM tip lies in the lead, we
would be in weak $\Gamma_{a}$. Then, the suppression of the Fano
resonance is expected to observe.


\section{Conclusion}

In the present paper we predicted an interesting feature of
non-Fermi liquid physics for the multichannel Kondo model, based
on the STM experimental setting (Fig.~\ref{fig1}). Non-Fermi
liquid physics often occurs at the quantum phase transition, and
presents challenges in both theoretical and experimental aspects,
for instance, quantum criticality in heavy fermion
materials.\cite{Review_HF} Heavy fermion materials can be modelled
by the Anderson lattice model. One heavy fermion quantum critical
point in the Anderson lattice model was argued to be captured in
the so called dynamical mean-field theoretical framework, more
concretely, the two impurity Anderson model with self-consistency,
expected to result in essentially similar physics with the
multichannel impurity model.\cite{DMFT_Two_Imp} As the first step
to understand non-Fermi liquid physics within the STM detection,
we employed the multichannel Anderson model for one source of the
non-Fermi liquid state.

We derived the Landauer-B\"uttiker formula for the tunneling
current from the STM tip to the multichannel impurity host based
on the Keldysh nonequilibrium formalism, where the tunneling
current is given by only the impurity Green function. Employing
the nonequilibrium NCA, we showed that the impurity Green function
exhibits universal power-law scaling at low energies. As a
consequence, the tunneling conductance turns out to exhibit weak
asymmetry but rather sharp cusp at zero energy, resulting from the
power-law scaling of the impurity Green function. The conventional
Fano resonance in Fermi liquids was shown to be suppressed. The
main prediction of our study is that the peak position in the
Fano-Kondo resonance does not shift, even increasing the tip
coupling constant, clearly distinguished from the Fermi liquid
theory. Quantum coherence of the impurity dynamics turns out to
play an important role in the Fano mechanism.

\acknowledgments

We would like to thank T. Takimoto for useful discussions. This
work was supported by the National Research Foundation of Korea
(NRF) grant funded by the Korea government (MEST) (No.
2009-0074542). One of the authors (M.-T.) was also supported by
the Vietnamese NAFOSTED.

\appendix

\section{Derivation of tunneling current}

In Appendix A we derive the tunneling current formula in
Eqs.~(\ref{currentfinal}) and (\ref{trans}). The lesser Green
functions in Eq.~(\ref{current}) are given by off diagonal
components of nonequilibrium Green functions, defined on the
Keldysh time contour which runs on the time axis from $- \infty$
to $\infty$, and goes back to $- \infty$,\cite{Keldysh,Meir}
\begin{eqnarray}
G^{c}_{d\sigma\tau,a\mathbf{k}\sigma}(t,t') &=& -i \big\langle
T_{c}
 d_{\sigma\tau}(t) a^{\dagger}_{\mathbf{k}\sigma}(t')
\big\rangle , \\
G^{c}_{c\mathbf{p}\sigma\tau,a\mathbf{k}\sigma}(t,t') &=& - i
\big\langle T_{c}
 c_{\mathbf{p}\sigma\tau}(t) a^{\dagger}_{\mathbf{k}\sigma}(t')
\big\rangle ,
\end{eqnarray}
where $T_c$ is the time ordering operator on the Keldysh time
contour. Differentiating the nonequilibrium Green functions with
respect to $t'$ and resorting to the Heisenberg equation of motion
\begin{eqnarray} && \frac{d a^{\dagger}_{\mathbf{k}\sigma}(t')}{d
t'} = - \frac{i}{\hbar} \big[H(t'),
a^{\dagger}_{\mathbf{k}\sigma}(t')\big] , \nonumber \end{eqnarray}
we obtain the following equations
\begin{eqnarray}
G^{c}_{d\sigma\tau,a\mathbf{k}\sigma}(t,t') &=& \int d t_1 \; V_a G^{c}_{d\sigma\tau,d\sigma\tau}(t,t_1)
g^{c}_{a\mathbf{k}}(t_1,t')  \nonumber \\
 + \sum_{\mathbf{p},\tau'} && \hspace{-0.7cm} \int d t_1 \;
t_c G^{c}_{d\sigma\tau,c\mathbf{p}\sigma\tau'}(t,t_1) g^{c}_{a\mathbf{k}}(t_1,t') ,
\label{gc1}\\
G^{c}_{d\sigma\tau,c\mathbf{p}\sigma\tau'}(t,t') &=& \delta_{\tau\tau'}
\int d t_1 \; V_c G^{c}_{d\sigma\tau,d\sigma\tau}(t,t_1)
g^{c}_{c\mathbf{p}}(t_1,t')  \nonumber \\
+ \sum_{\mathbf{k}} && \hspace{-0.7cm} \int d t_1 \;
t_c^* G^{c}_{d\sigma\tau,a\mathbf{k}\sigma}(t,t_1) g^{c}_{c\mathbf{p}}(t_1,t') , \label{gc2}
\end{eqnarray}
where $g^{c}_{a\mathbf{k}}(t,t')$, $g^{c}_{c\mathbf{k}}(t,t')$ are
the nonequilibrium Green functions for the isolated noninteracting
conduction electrons in the tip and host, respectively.

Inserting Eq.~(\ref{gc2}) into Eq.~(\ref{gc1}), we obtain
\begin{eqnarray}
\lefteqn{
G^{c}_{d\sigma\tau,a\sigma}(t,t') = \int \!\!d t_1 \; V_a G^{c}_{d\sigma\tau,d\sigma\tau}(t,t_1)
g^{c}_{a}(t_1,t') }  \nonumber \\
 && \hspace{-0.7cm}+ \int \!\! d t_1 \!\!\!\int \!\! d t_2 \;
t_c V_c G^{c}_{d\sigma\tau,d\sigma\tau}(t,t_1) g^{c}_{c}(t_1,t_2)  g^{c}_{a}(t_2,t')  \nonumber \\
&& \hspace{-0.7cm} +  \int \!\!d t_1 \!\!\!\int \!\!d t_2\;
M |t_c|^2  G^{c}_{d\sigma\tau,a\sigma}(t,t_1) g^{c}_{c}(t_1,t_2) g^{c}_{a}(t_2,t') , \label{gc2c}
\end{eqnarray}
where
$g^{c}_{a(c)}(t,t')=\sum_{\mathbf{k}}g^{c}_{a(c)\mathbf{k}}(t,t')$.
Using the Langreth's rule of analytical continuation on the real
time axis,\cite{Langreth,Jauho} we can obtain the retarded
(advanced) and lesser (or greater) Green functions from
Eq.~(\ref{gc2c})
\begin{eqnarray}
\lefteqn{
G^{R/A}_{d\sigma\tau,a\sigma} =  V_a G^{R/A}_{d\sigma\tau,d\sigma\tau} \ast g^{R/A}_{a}
} \nonumber \\
&& \hspace{1.5cm} + t_c V_c G^{R/A}_{d\sigma\tau,d\sigma\tau}\ast g^{R/A}_{c} \ast  g^{R/A}_{a} \nonumber \\
&& \hspace{1.5cm} +
M |t_c|^2  G^{R/A}_{d\sigma\tau,a\sigma} \ast g^{R/A}_{c} \ast g^{R/A}_{a} ,  \label{agc1}\\
\lefteqn{ G^{<}_{d\sigma\tau,a\sigma} =  V_a \Big( G^{R}_{d\sigma\tau,d\sigma\tau} \ast
g^{<}_{a} + G^{<}_{d\sigma\tau,d\sigma\tau} \ast g^{A}_{a}  \Big) }  \nonumber \\
&& +
t_c V_c \Big( G^{R}_{d\sigma\tau,d\sigma\tau} \ast g^{R}_{c} \ast  g^{<}_{a}
+ G^{R}_{d\sigma\tau,d\sigma\tau} \ast g^{<}_{c} \ast  g^{A}_{a}  \nonumber \\
&& \hspace{1.5cm}+ G^{<}_{d\sigma\tau,d\sigma\tau} \ast g^{A}_{c} \ast  g^{A}_{a} \Big)
  \nonumber \\
&& +
M |t_c|^2 \Big( G^{R}_{d\sigma\tau,a\sigma} \ast g^{R}_{c} \ast g^{<}_{a} +
G^{R}_{d\sigma\tau,a\sigma} \ast g^{<}_{c} \ast g^{A}_{a}   \nonumber \\
&& \hspace{1.5cm} +G^{<}_{d\sigma\tau,a\sigma} \ast g^{A}_{c} \ast g^{A}_{a} \Big),
\label{agc2}
\end{eqnarray}
where the superscripts, $R$, $A$, $<$, denote the retarded,
advanced, lesser Green functions, respectively. For simplifying to
write equations, we use the so called $\ast$ notation, defined as
$A(t)\ast B(t') = \int d t_{1} A(t,t_1) B(t_1,t')$.

In the steady state Eqs.~(\ref{agc1})-(\ref{agc2}) are easily
solved, making the Fourier transformation. As a result, we find
\begin{eqnarray}
\lefteqn{
G^{<}_{d\sigma\tau,a\sigma}(\omega) = G^{R}_{d\sigma\tau,d\sigma\tau}(\omega) Z^{A}(\omega)
\Big[ V_a g^{<}_{a}(\omega) + } \nonumber \\
&& \hspace{-0.5cm} \big( t_c V_c+M |t_c|^2 \big) \big( g^{R}_{c}(\omega) g^{<}_{a}(\omega)+
g^{<}_{c}(\omega) g^{A}_{a}(\omega)  \big)
\Big] + \nonumber \\
&& \hspace{-0.5cm} G^{<}_{d\sigma\tau,d\sigma\tau}(\omega) Z^{A}(\omega)
\Big[ V_a g^{A}_{a}(\omega) + t_c V_c  g^{A}_{c}(\omega) g^{A}_{a}(\omega) \Big],
\label{lesser1}
\end{eqnarray}
where $Z^{R/A}(\omega) =1/[1 - M |t_c|^2 g^{R/A}_{c}(\omega) g^{R/A}_{a}(\omega)] $.

Performing in a similar way, we can express the Green function
$G^{c}_{c\sigma,a\sigma}(t,t')=\sum_{\mathbf{k},\mathbf{p},\tau}
G^{c}_{c\mathbf{k}\sigma\tau,a\mathbf{p}\sigma}(t,t')$ with
$G^{c}_{d\sigma,a\sigma}(t,t')=\sum_{\tau}G^{c}_{d\sigma\tau,a\sigma}(t,t')$.
One can verify
\begin{eqnarray}
\lefteqn{
G^{<}_{c\sigma,a\sigma}(\omega) = M t_c Z^{R}(\omega)
\big[ g^{R}_c(\omega) g^{<}_a(\omega) + g^{<}_a(\omega) g^{A}_a(\omega) \big] } \nonumber \\
&& \hspace{2cm} \big[ 1 + M |t_c|^2 Z^A(\omega) g^{A}_c(\omega) g^{A}_a(\omega) \big]   \nonumber \\
&& + G^{<}_{d\sigma,a\sigma}(\omega) Z^{R}(\omega)
\big[ V_c^{*} g^{R}_c(\omega) + M t_c V^{*}_a g^{R}_c(\omega) g^{R}_a(\omega) \big] \nonumber \\
&&+  G^{A}_{d\sigma,a\sigma}(\omega) Z^{R}(\omega) \big[ V^{*}_c g^{<}_c(\omega)
+ ( g^{R}_c(\omega) g^{<}_a(\omega) +  \nonumber \\
&& g^{<}_c(\omega) g^{A}_a(\omega) )
( M t_c V^*_a + M |t_c|^2 Z^A(\omega) \nonumber \\
&& (V^{*}_c g^{A}_c + M t_c V^{*}_a g^{A}_c(\omega) g^{A}_a(\omega))
\big] .
\label{lesser2}
\end{eqnarray}

Using Eqs.~(\ref{lesser1})-(\ref{lesser2}), we can express the
steady current in Eq.~(\ref{current}) with the nonequilibrium
Green functions of the impurity. For simplicity, we will consider
a flat density of states for the tip and host conduction electrons
in the wide band limit, given by
\begin{eqnarray}
g^{R/A}_{a(c)}(\omega) &=& \mp i \pi \rho_{a(c)} , \\
g^{<}_{a(c)}(\omega) &=& 2 \pi i \rho_{a(c)} f_{a(c)}(\omega),
\end{eqnarray}
where $\rho_{a(c)}$ is the density of states for noninteracting
tip (host) conduction electrons at the Fermi level, and
$f_{a(c)}(\omega)$ is its Fermi-Dirac distribution function. From
Eqs.~(\ref{current}), (\ref{lesser1}), and (\ref{lesser2}) we
obtain
\begin{eqnarray}
\lefteqn{
J_{t\rightarrow h} = \frac{e}{\hbar} \sum_{\sigma\tau}  \int \frac{d\omega}{2\pi} \Big\{
\frac{T_0}{2}\big( f_a(\omega)-f_c(\omega)\big) } \nonumber \\
&& +  G^{<}_{d\sigma\tau,d\sigma\tau}(\omega) T_{ac1} + G^{R}_{d\sigma\tau,d\sigma\tau}(\omega) 2 T_{ac1} f_{c}(\omega)
\nonumber \\
&&+ G^{R}_{d\sigma\tau,d\sigma\tau}(\omega) T_{ac2} \big( f_a(\omega)-f_c(\omega)\big)
+{\rm h.c.}\Big\}, \label{currentth}
\end{eqnarray}
where
\begin{eqnarray*}
T_{0} &=& \frac{4 \gamma}{(1+M \gamma)^2} , \\
T_{ac1} &=& i \pi \rho_a \frac{|V_a+iV_c t_c \pi \rho_c|^2}{(1+M \gamma)^2} , \\
T_{ac2} &=& 2 i \pi \rho_a   \frac{(1-M\gamma)(V_a-i V_c t_c \pi \rho_c)^2}{(1+M \gamma)^3} .
\end{eqnarray*}
Here $\gamma=\pi^2 |t_c|^2 \rho_a \rho_c$ is a measure of the
strength for the direct tunneling of conduction electrons between
the tip and host.

In a similar way we can find a current flowing from the host to
the tip based on Eq. (\ref{currentht}),
\begin{eqnarray}
\lefteqn{
J_{h\rightarrow t} = \frac{e}{\hbar}  \sum_{\sigma\tau}  \int \frac{d\omega}{2\pi} \Big\{
\frac{T_0}{2}\big( f_c(\omega)-f_a(\omega)\big) } \nonumber \\
&& + G^{<}_{d\sigma\tau,d\sigma\tau}(\omega) T_{ca1}
+ G^{R}_{d\sigma\tau,d\sigma\tau}(\omega) 2 T_{ca1} f_{a}(\omega)
\nonumber \\
&&+ G^{R}_{d\sigma\tau,d\sigma\tau}(\omega) T_{ca2} \big( f_c(\omega)-f_a(\omega)\big)
+{\rm h.c.}\Big\} \nonumber \\
&& + \frac{e}{\hbar}   \frac{(M-1) \gamma}{1+M\gamma} \sum_{\sigma\tau}  \int \frac{d\omega}{2\pi} \Big\{
 G^{<}_{d\sigma\tau,d\sigma\tau}(\omega) (\Delta T_c + \Delta T_a) \nonumber \\
&& + 2 G^{R}_{d\sigma\tau,d\sigma\tau}(\omega) \big[
\Delta T_c f_c(\omega) + \Delta T_a f_a(\omega) \nonumber \\
&& + \Delta T (f_c(\omega)-f_a(\omega)) \big] + {\rm h.c.} \Big\}
, \label{currenthta}
\end{eqnarray}
where
\begin{eqnarray*}
\Delta T_a &=& i \Gamma_a \frac{1}{1+M\gamma} ,\\
 \Delta T_c &=& i \Gamma_c \Big(1+\frac{1}{1+M\gamma} \Big)  , \\
\Delta T &=& 2 i \Gamma_c \frac{1}{(1+M\gamma)^2} + i \Gamma_a \frac{M\gamma-1}{(1+M\gamma)^2}  .
\end{eqnarray*}
$T_{ca1}$, $T_{ca2}$ are just $T_{ac1}$, $T_{ac2}$, changing
indices $a \leftrightarrow c$ while the hopping $t_c$ is
unchanged.

One can notice that the first term of the current $J_{h\rightarrow
t}$ in Eq.~(\ref{currenthta}) is just the current $J_{t\rightarrow
h}$ in Eq.~(\ref{currentth}) if the tip and host are interchanged
with each other. In the single channel case ($M=1$) the second
term of the current $J_{h\rightarrow t}$ in Eq.~(\ref{currenthta})
vanishes, thus the current formula satisfies the symmetry between
the tip and host. However, in the multichannel case ($M>1$) the
symmetry is broken, because host conduction electrons are
multichannel, whereas tip-conduction electrons are single-channel.

\section{Solution for NCA equations}

In Appendix B we derive the NCA solution Eq.~(\ref{yfs}) with
Eq.~(\ref{scaling}) and Eqs.~(\ref{ls1}), (\ref{ls2}) from
Eqs.~(\ref{desf}), (\ref{desb}) and Eqs.~(\ref{les1}),
(\ref{les2}), respectively. Introducing inverse Green
functions,\cite{Muller,Bickers}
\begin{eqnarray}
Y_f(\omega) &=& -\big[G_f^R(\omega)\big]^{-1} , \\
Y_b(\omega) &=& -\big[G_b^R(\omega)\big]^{-1} ,
\end{eqnarray}
one can rewrite Eqs. (\ref{desf}) and (\ref{desb}) as
\begin{eqnarray}
\frac{d Y_{f}(\omega)}{d\omega} &=& -1 - \frac{M \Gamma_c}{\pi} Y^{-1}_b(\omega), \label{y1} \\
\frac{d Y_{b}(\omega)}{d\omega} &=& -1 - \frac{N \Gamma_c}{\pi}
Y^{-1}_f(\omega) . \label{y2}
\end{eqnarray}
Then, we find the exact relation between $Y_{f}(\omega)$ and
$Y_b(\omega)$
\begin{equation}
\frac{\pi}{M \Gamma_c} Y_b \exp\Big(\frac{\pi}{M \Gamma_c} Y_b
\Big) = \Big( \frac{Y_f}{T_K} \Big)^{N/M} \exp\Big(\frac{\pi}{M
\Gamma_c} Y_f \Big) , \label{fb}
\end{equation}
where $T_K = D [ M \Gamma_c / \pi D ]^{M/N} \exp[\pi
\varepsilon_d/N \Gamma_c]$ is the Kondo energy scale.

Solving Eq.~(\ref{fb}), we obtain
\begin{equation}
Y_b = \frac{M \Gamma_c}{\pi} \; W\Big[\Big( \frac{Y_f}{T_K}
\Big)^{N/M} \exp\Big(\frac{\pi T_K}{M \Gamma_c} \frac{Y_f}{T_K}
\Big) \Big] , \label{bviaf}
\end{equation}
where $W[x]$ is the Lambert $W$ function, given by $x=W[x]
\exp(W[x])$.\cite{Corless} Inserting $Y_b$ in Eq.~(\ref{bviaf})
into Eq.~(\ref{y1}), we find the solution given by Eq. (\ref{yfs})
with Eq. (\ref{scaling}).

Although the NCA equations (\ref{les1})-(\ref{les2}) for lesser
self-energies are identical to the differential equations
(\ref{desf})-(\ref{desb}) of retarded self-energies, the Dyson
equations (\ref{d3})-(\ref{d4}) for the lesser Green functions
have a different structure. Together with the Dyson equations
(\ref{d3})-(\ref{d4}), Eqs.~(\ref{les1})-(\ref{les2}) can be
rewritten as
\begin{eqnarray}
\frac{d [ F^{<}(\omega) Y_{f}^2(\omega)]}{d\omega} &=& \frac{M \Gamma_c}{\pi} B^{<}(\omega), \label{ll1} \\
\frac{d [B^{<}(\omega) Y_{b}^2(\omega) ]}{d\omega} &=& \frac{N
\Gamma_c}{\pi} F^{<}(\omega) , \label{ll2}
\end{eqnarray}
for frequency below $E_0$. Then, we find Eqs. (\ref{ls1}) and
(\ref{ls2}) as their solution.

\section{Effective channel couplings}

In this Appendix we calculate the hybridization coupling of the
channel $\tau=1$ $\Gamma^{c(1)}$ for the effective action of
Eq.~(\ref{effaction}).

We start from the following matrix identity \begin{equation}
(\mathbf{R}+x \mathbf{U})^{-1} = \mathbf{R}^{-1} - \frac{x}{1+x S}
 \mathbf{F} , \label{Matrix_Identity}
\end{equation}
where $\mathbf{R}_{ij}=R_{i} \delta_{ij}$ is a diagonal matrix,
$\mathbf{U}_{ij} = 1$ for all $i$, $j$, $x$ is a number, $S =
\sum_{i} {R}_{i}^{-1}$, and $\mathbf{F}_{ij} =
{R}_{i}^{-1}{R}_{j}^{-1}$. Taking $\mathbf{R} = \delta_{kk'}
(\omega-\varepsilon_{k})$ and $x = - M |t_c|^2 g_{a}(\omega)$, we
find the inversion of $[g^{c(1)}_{kk'}(\omega)]^{-1}$,
\begin{equation}
g^{c(1)}_{kk'}(\omega) =
\frac{\delta_{kk'}}{\omega-\varepsilon_{k}} + \frac{M |t_c|^2
g_{a}(\omega)}{1-M |t_c|^2 g_{a}(\omega) g_{c}(\omega)}
\frac{1}{\omega-\varepsilon_{k}}
\frac{1}{\omega-\varepsilon_{k'}},
\end{equation}
where $g_{c}(\omega)=\sum_{\mathbf{k}}
1/(\omega-\varepsilon_{k})$. Performing the summation of
$\mathbf{k}$, $\mathbf{k'}$ we obtain
\begin{eqnarray}
\sum_{\mathbf{k}\mathbf{k}'} g^{c(1)}_{kk'}(\omega) &=&
g_{c}(\omega) + \frac{M |t_c|^2 g_{a}(\omega)}{1-M |t_c|^2
g_{a}(\omega) g_{c}(\omega)} \big(g_{c}(\omega)\big)^2 \nonumber
\\
&=& \frac{g_{c}(\omega)}{1-M |t_c|^2 g_{a}(\omega) g_{c}(\omega)}
. \label{ag}
\end{eqnarray}
In the wide band limit of a constant density of states for
conduction electrons in both the host and tip, Eq.~(\ref{ag})
results in
\begin{equation}
\Gamma^{c(1)} = \frac{\Gamma_c}{1+M\gamma} .
\end{equation}
It also shows $\Gamma^{c(1)} \leq \Gamma_{c}$.

We prove the identity Eq. (\ref{Matrix_Identity}). Consider
\begin{equation}
(\mathbf{R}+x \mathbf{U})^{-1} = \sum_{n=0}^{\infty} (-1)^{n}
x^{n} \big( \mathbf{R}^{-1} \mathbf{U}\big)^{n} \mathbf{R}^{-1} .
\label{matrix0}
\end{equation}
We introduce the following identity
\begin{eqnarray}
\lefteqn{ \sum_{n=1}^{\infty} (-1)^{n} x^{n} \big( \mathbf{R}^{-1}
\mathbf{U}\big)^{n} \mathbf{R}^{-1}= } \nonumber \\
&& -x \Big[ \sum_{n=0}^{\infty} (-1)^{n} x^{n} \big(
\mathbf{R}^{-1} \mathbf{U}\big)^{n+1} \mathbf{R}^{-1} \Big]
 . \label{matrixi}
\end{eqnarray}
Since the matrix $\mathbf{R}$ is diagonal, we can write its
inverse as $(\mathbf{R}^{-1} )_{ij} = \tilde{R}_{i} \delta_{ij}$,
where $\tilde{R}_{i}=1/R_i$. Calling  $(\mathbf{R}^{-1}
\mathbf{U})_{ij} = \tilde{R}_{i}$, where elements in each raw are
identical, we find \begin{eqnarray} \big[( \mathbf{R}^{-1}
\mathbf{U})^{n} \big]_{ij} = \tilde{R}_{i} S^{n-1} \label{matrixs}
\end{eqnarray}
with $S=\sum_{i} \tilde{R}_{i}$. Inserting Eq.~(\ref{matrixs})
into Eq.~(\ref{matrixi}), we obtain
\begin{eqnarray}
\sum_{n=1}^{\infty} (-1)^{n} x^{n} \big( \mathbf{R}^{-1}
\mathbf{U}\big)^{n} \mathbf{R}^{-1} &=& -x \Big[
\sum_{n=0}^{\infty} (-1)^{n}
x^{n} S^{n} \mathbf{F} \Big]  \nonumber \\
&=& - \frac{x}{1+x S} \mathbf{F} ,  \label{matrixz}
\end{eqnarray}
where $\mathbf{F}_{ij}=\tilde{R}_{i} \tilde{R}_{j}$. Resorting to
Eqs.~(\ref{matrix0}) and (\ref{matrixz}), finally,  we reach Eq.
(\ref{Matrix_Identity}).

\end{document}